\documentclass{article}
\pdfoutput=1
\usepackage{times}
\usepackage{graphicx} % more modern
\usepackage{float}
\usepackage{subfigure} 
\usepackage{natbib}
\usepackage{hyperref}

\usepackage[accepted]{grcon2016}

\grcontitlerunning{Implementation of FGPA based Channel Sounder for Large scale antenna systems using RFNoC on USRP Platform}

\begin{document}
\twocolumn[
\grcontitle{Implementation of FGPA based Channel Sounder for Large scale antenna systems using RFNoc on USRP Platform}

\grconauthor{Bhargav Gokalgandhi}{bvg8@scarletmail.rutgers.edu}
\grconaddress{WINLAB, Rutgers University, New Jersey}
\grconauthor{Prasanthi Maddala}{prasanti@winlab.rutgers.edu}
\grconaddress{WINLAB, Rutgers University, New Jersey}
\grconauthor{Ivan Seskar}{seskar@winlab.rutgers.edu}
\grconaddress{WINLAB, Rutgers University, New Jersey}

\grconkeywords{software radio, gnu radio, dsp, RFNoC, USRP, channel sounding, GRCON}
]

\begin{abstract}

This paper concentrates on building a multi-antenna FPGA based Channel Sounder with single transmitter and multiple receivers to realize wireless propagation characteristics of an indoor environment. A DSSS signal (spread with a real maximum length PN sequence) is transmitted, which is correlated with the same PN sequence at each receiver to obtain the power delay profile . Multiple power delay profiles are averaged and the result is then sent to host. To utilize high bandwidth, the computationally expensive tasks related to generation and parallel correlation of PN sequences are moved to the FPGA present in each USRP (Universal Software Radio Peripheral). Channel sounder blocks were built using Vivado HLS and integrated with RFNoC (RF Network on Chip) framework, which were then used on USRP X310 devices.

\end{abstract}

\section{Introduction}\label{intro}

Channel sounding is the process of evaluating the characteristics of a radio environment. This is essential for developing a statistical channel model for simulations. Knowledge of channel characteristics can also help in designing  better antenna systems which results in communication systems with increased throughput and higher reliability. 

For 5G communication systems, where ultra-wide bandwidths (mmWave) and high number of antennas (Massive MIMO) are to be used, Channel State Information (CSI) acquisition becomes extremely complex. For such systems, using traditional CSI acquisition algorithms can place significant overhead on the system, causing a loss in spectral efficiency . For example, in systems with pilot-based CSI acquisition, the number of pilot signals to be used increases with the number of antennas and the bandwidth used. Processing these pilot signals is computationally intensive and time consuming. However, studying the radio environment, obtaining prior statistical information, and developing models, can result in designing simpler algorithms specific to the environment, resulting in an efficient communication system. 

The field of channel sounding has been an area of research for a long time. Modeling of wireless channel characteristics using statistical and propagation data gained though channel sounding can be seen in papers like  \cite{german2001wireless},\cite{ritcher2003maximum},\cite{thoma2004rimax},\cite{richter2005estimation},\cite{ciccognani2005time},\cite{thoma2005multidimensional}.\cite{dezfooliyan2012evaluation} talks about creating a spread spectrum based UWB channel sounding system. Using spread spectrum based methods to create UWB channel sounding systems can be seen in \cite{haese1999high},\cite{merwaday2014usrp},\cite{le2013spread},\cite{islam2013wireless},\cite{thoma2001mimo},\cite{pirkl2008optimal}. \cite{merwaday2014usrp},\cite{le2013spread},\cite{islam2013wireless} have used USRP platform for creating channel sounding systems while \cite{thoma2001mimo},\cite{kmec2005novel},\cite{kolmonen2010dynamic},\cite{maharaj2005cost},\cite{elofsson2016software} concentrate on channel sounder architectures for MIMO systems with \cite{elofsson2016software} using USRPs for channel sounder implementation. Some examples of channel measurements made using MIMO channel sounder can be seen in \cite{kim2015large},\cite{liu2016mimo},\cite{salous2016wideband}. \cite{salous2016wideband} and \cite{panda2012fpga} show the FPGA implementation of LFSR based PN sequences. 

The channel sounders constructed and implemented in the above papers using USRPs or any other SDRs use less number of antennas or low bandwidth. That is because the above approaches have higher host side processing which requires all the data received on every antenna to be down-converted and sent to the host. The number of samples to be sent to the host for processing increases with increasing number of antennas as well as increasing bandwidth which in turn restricts the maximum sampling rate that can be used. Because of that, implementation of MIMO channel sounders requires specific hardware platforms to be built which can support parallel processing. \cite{elofsson2016software} shows how using FPGA for channel sounding implementation can help but the paper is limited to implementation of $2\times 2$ MIMO.

This paper takes into consideration the problems caused due to simultaneous increase in number of antennas and bandwidth, and creates a real-time UWB channel sounder for large scale antenna systems. We present an implementation of FPGA and SDR based UWB Channel sounder for Large scale Antenna systems and show the results obtained when the sounder is used for a 16 antenna system at 100 MHz bandwidth.

This paper is organized as follows. Section \ref{impl} discusses FPGA implementation of spread spectrum channel sounder, by giving details  about the different blocks implemented and integrated with RFNoC. Experimental setup is described in Section \ref{setup}, measurements and results are presented in Section \ref{res}. The paper is then concluded in Section \ref{conc}.

\section{FPGA implementation of channel sounder}\label{impl}
\begin{figure}[ht]
  \begin{center}
    \centerline{\includegraphics[width=\columnwidth]{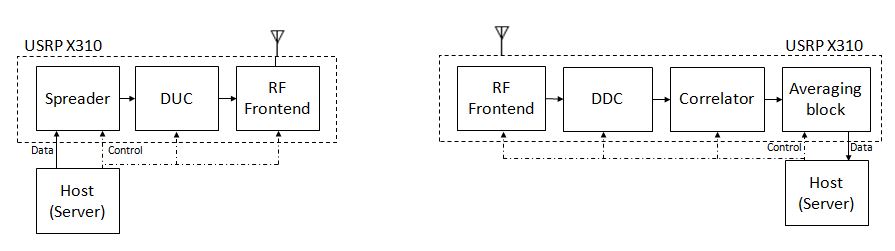}}
    \caption{USRP X310 Spread spectrum Channel Sounder}
    \label{channel_sounder_block_diagram}
  \end{center}
\end{figure}

A spread spectrum channel sounder using USRPs can be built as shown in Figure~\ref{channel_sounder_block_diagram}. Spectrum spreader modulates data samples with a maximal length PN sequence. Modulated samples are passed through a shaping filter, upconverted to the desired carrier frequency and transmitted over the air. Received baseband samples are correlated with the conjugate of transmit PN sequence, to identify delayed versions of the transmitted PN sequence, thus resolving the multipath components of the channel. Correlation power or power delay profile obtained is averaged over multiple data symbols to eliminate spurious correlation peaks due to noise~\cite{gan2005path}.
While building the above channel sounder, spectrum spreader on the transmit side, and correlator, averaging block on the receive side are the only blocks that needed to be implemented, as shaping filters (included in DDC, DUC) and radio interface blocks are already a part of the default USRP FPGA image. Instead of using software modules for the above 3 blocks, we have chosen to implement them in the FPGA,  to  enable multi-channel, high bandwidth, real time channel sounding. These blocks were then integrated with RFNoC framework as described in the following section. Though the blocks were built to be used in this USRP channel sounding system, with a standard AXI stream interface, they can be easily ported to any application or platform. Each of the 3 blocks implemented is described in the subsequent sections.

\subsection{RFNoC channel sounder}
\begin{figure}[ht]
  \begin{center}
    \centerline{\includegraphics[width=\columnwidth]{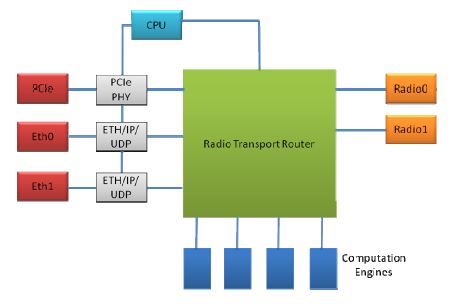}}
    \caption{RFNoC Architecture}
    \label{rfnoc_arch}
  \end{center}
\end{figure}
RFNoC (RF Network on Chip) is a unique processing and routing architecture, developed by Ettus Research to reduce the FPGA development time on Third Generation Ettus Research USRPs~\cite{malsbury2013simplifying}. This architecture provides basic functionality for the SDR, such as host communication, packet processing, radio interfaces, clocking etc., while allowing users to easily integrate custom IP as "Computation Engines".
 
Figure~\ref{rfnoc_arch} shows how various blocks in an RFNoC design are connected to the Radio transport router(RTR), which acts as a cross switch for routing data/control packets between them. Once an RFNoC build is made with the required blocks, connections between them or the flow graph can be established from the host computer by sending a few UHD (USRP Hardware driver) commands to the Radio transport router.

\subsubsection{RFNoC DSSS Transmitter}
\begin{figure}[ht]
  \begin{center}
    \centerline{\includegraphics[width=\columnwidth]{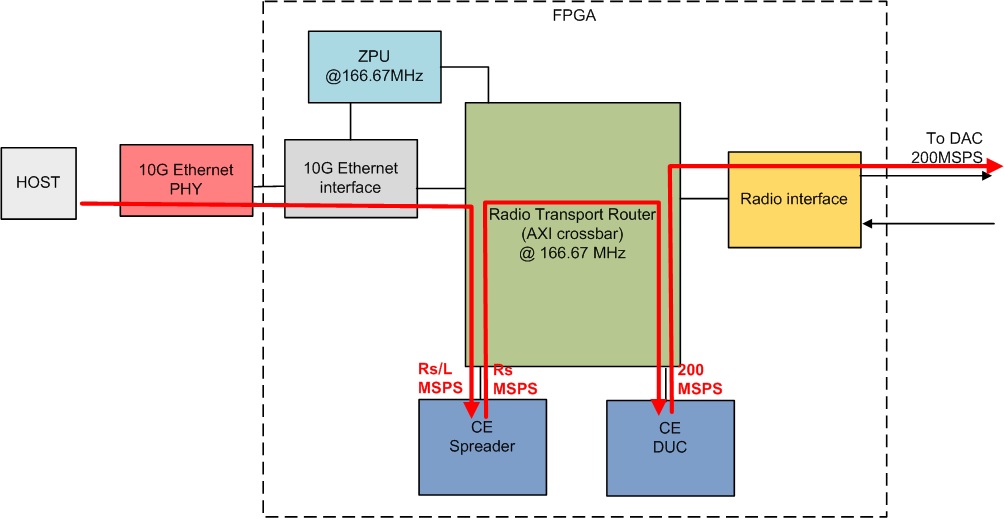}}
    \caption{RFNoC DSSS Transmitter showing transmit data flow }
    \label{rfnoc_tx}
  \end{center}
\end{figure}
DSSS transmitter as a part of the channel sounding system was built by adding a spectrum spreader computation engine(CE) to the RFNoC framework. A DUC CE provided by Ettus Research was also used in the transmitter. As shown is Figure~\ref{rfnoc_tx} the RTR was configured to build a flow graph to route data from the host to the spectrum spreader. The spreader gives out samples at rate Rs, which is the channel sounding rate, which corresponds to the bandwidth being sounded. These samples are then routed to the DDC which interpolates them to 200MSPS and sends them out to a Radio.

\subsubsection{RFNoC Channel sounding Receiver}
\begin{figure}[ht]
  \begin{center}
    \centerline{\includegraphics[width=\columnwidth]{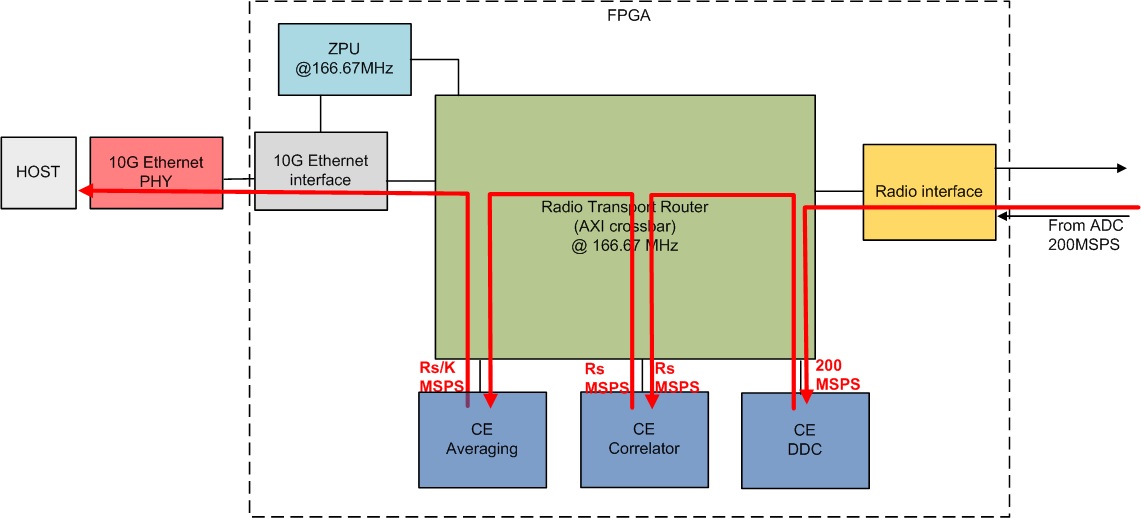}}
    \caption{RFNoC Channel sounding receiver showing receive data flow }
    \label{rfnoc_rx}
  \end{center}
\end{figure}
Channel sounding receiver was similarly built by adding correlator, averaging CEs to the RFNoC framework. A DDC CE provided by Ettus Research was also used. Samples received from a Radio at a rate of 200MSPS are routed to DDC, which decimates them to rate Rs. 

These samples pass through the correlator and averaging CEs before being sent to the host, as shown in Figure~\ref{rfnoc_rx}. Averaging CE further reduces the data rate by a factor of K, the averaging factor.

\subsection{Spectrum spreader}
\begin{figure}[ht]
  \begin{center}
    \centerline{\includegraphics[width=\columnwidth]{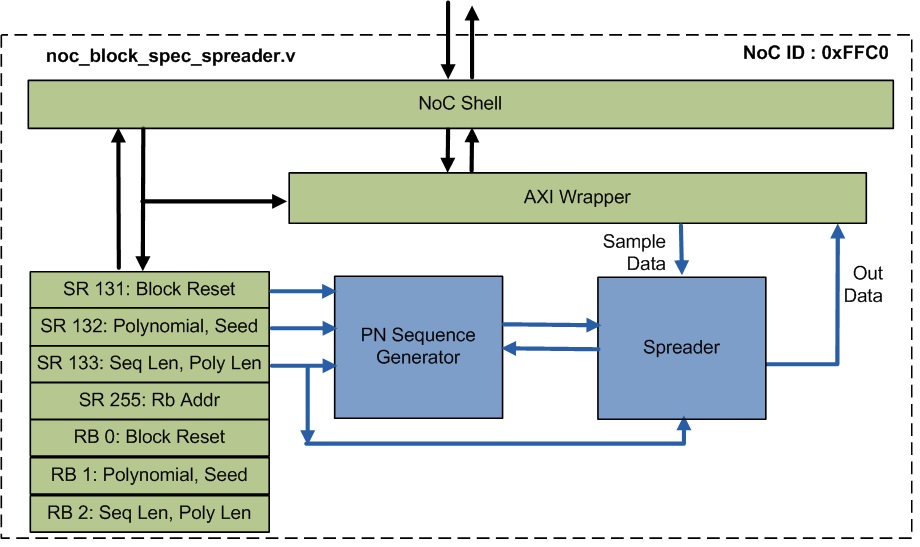}}
    \caption{Spectrum Spreader Computation Engine}
    \label{spreader_ce}
  \end{center}
\end{figure}

Spectrum spreader CE takes in data symbols in SC16 format(signed complex numbers with 16 bit real and imaginary parts) and spreads them using a real PN sequence giving L SC16 output symbols for each input, where L is the sequence length. The PN sequence is locally generated using a user-provided generator polynomial and seed. Setting registers~\cite{pendlum2014rfnocfpga} as shown in Figure~\ref{spreader_ce} allow users to provide the generator polynomial, seed and sequence length.. 

\begin{figure}[H]
  \begin{center}
    \centerline{\includegraphics[width=\columnwidth]{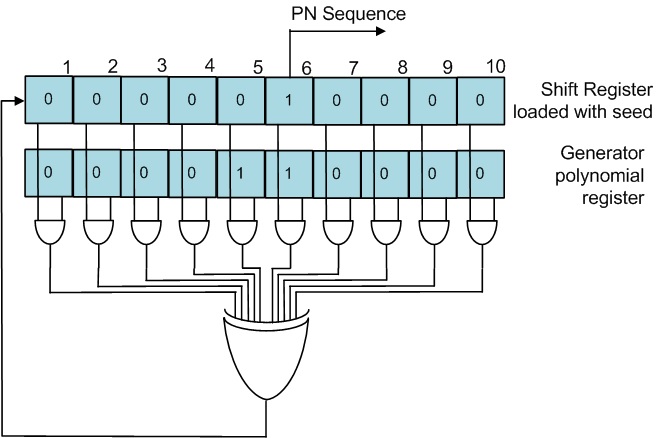}}
    \caption{PN sequence generator programmed with an order 6 generator polynomial}
    \label{PNseqgen}
  \end{center}
\end{figure}

LFSR based programmable PN sequence generator as shown in Figure~\ref{PNseqgen} was implemented. It can take a generator polynomial up to an order of 10, i.e., the longest sequence that it can generate is of length 1023. For a polynomial of order N, the output is taken from the Nth bit.

Maximal length sequences which have excellent auto correlation properties are used for channel sounding. Other sequences could be generated and used when the spectrum spreader block is used in a different application. Vivado HLS was used to generate RTL code for the PN sequence generator and spreader blocks.

\subsection{Correlator}
\begin{figure}[h]
  \begin{center}
    \centerline{\includegraphics[width=\columnwidth]{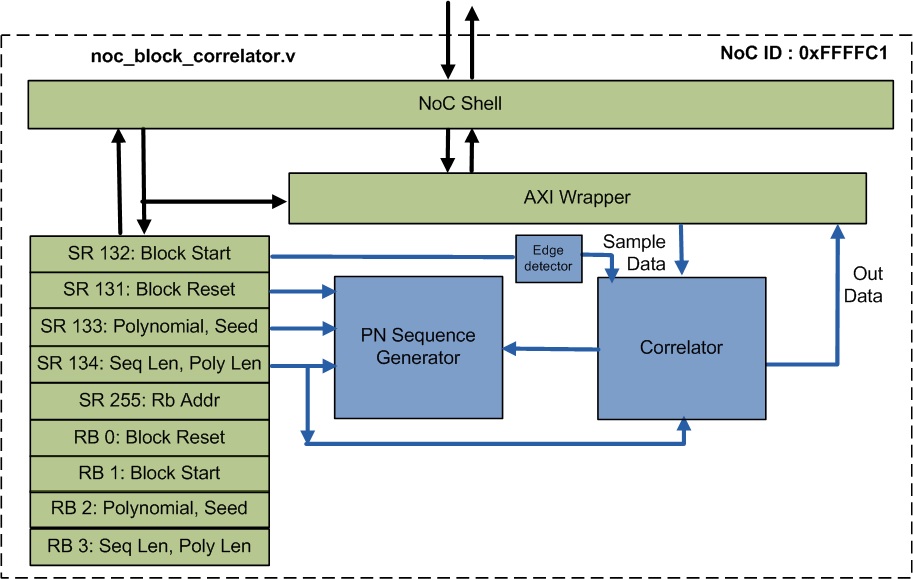}}
    \caption{Correlator Computation Engine}
    \label{correlator_ce}
  \end{center}
\end{figure}

Correlator CE starts processing incoming samples when it receives a start pulse from the user. For each SC16 input sample, it gives out a 32 bit unsigned integer, the correlation power. A parallel correlator was used to give out results at the same rate as the incoming sample rate, enabling the usage of maximum bandwidth possible. Setting registers as shown in Figure~\ref{correlator_ce} allow users to program the correlator to be used with a desired sequence. PN sequence generator described in the previous section was used in the correlator CE as well.

\begin{figure}[ht]
  \begin{center}    \centerline{\includegraphics[width=\columnwidth]{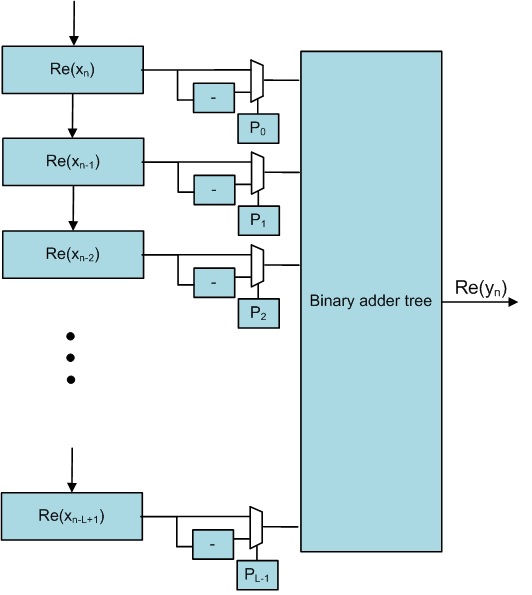}}
    \caption{Parallel Correlator computing real part of correlation}
    \label{pcorr}
  \end{center}
\end{figure}

Two parallel correlator structures as shown in Figure~\ref{pcorr} were used for real and imaginary parts of the incoming samples. Samples are pushed into the shift register  storage which can hold up to 512 samples. So the correlator can be used with sequences up to length 512. For a sequence of length L, the \(n^{th}\) correlator output can be given as
\begin{equation}  
    y_n = \sum\limits_{l=0}^{L-1}p_lx_{n-l}
    \label{corr_eq}  
\end{equation}

Each time a new sample comes in, the oldest sample is pushed out as it is no longer required. Since the PN sequence is real and binary, multiplication in Equation~\ref{corr_eq} comes down to selecting the original sample when the coefficient is +1 and 2's complement of the sample when it is -1~\cite{garrett1999low}. Correlation thus obtained is a complex number, the squared magnitude of which gives the correlation power. 

\subsection{Averaging Block}
\begin{figure}[h]
  \begin{center}
    \centerline{\includegraphics[width=\columnwidth]{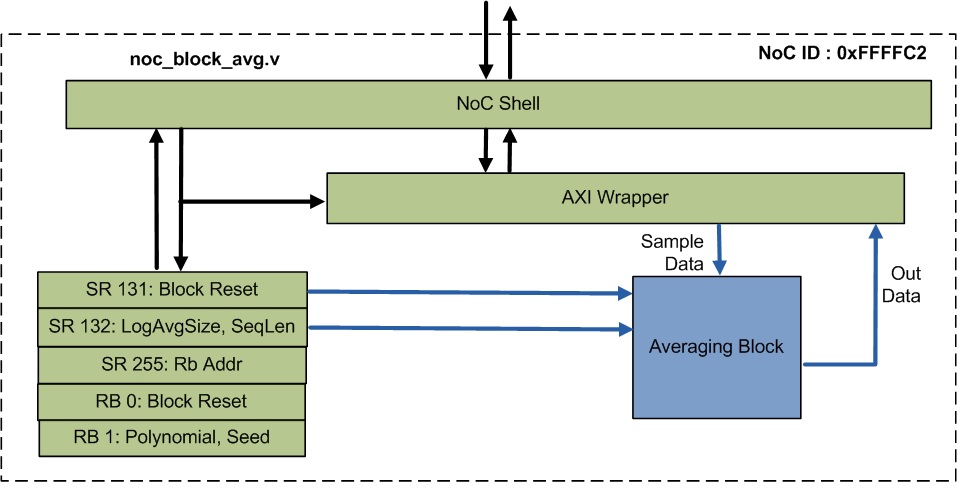}}
    \caption{Averaging Computation Engine}
    \label{averaging_ce}
  \end{center}
\end{figure}

Averaging CE is used to compute the average of K incoming packets. It is used in this channel sounding system to eliminate spurious peaks that might appear in the power delay profile due to noise. Also, since it reduces the data rate by a factor of K (where K is the averaging size), it enables the user to collect data from large number of antennas without exhausting the Ethernet link between USRPs and host. Setting registers as shown in Figure~\ref{averaging_ce} allow users to set the averaging factor.Only averaging factors \(<= 128\) that are a power of 2 are taken, so that division operation reduces to a right shift operation. 
\subsection{FPGA Utilization} \label{fpgautil}
{\scriptsize
\begin{table}[h]
\begin{center}
\begin{tabular}{|p{2cm}|c|c|c|}
\hline
& LUT & FF & BRAM  \\
\hline
Available in XC7K410T & 254200 & 508400 & 795 \\
\hline
\hline
& \%LUT & \%FF & \%BRAM \\
\hline
Spreader CE & 1 & 0.66 & 0.88 \\
\hline
Correlator CE 256 & 9.4 & 7.35 & 0.75 \\
\hline
Correlator CE 512 & 17.9 & 14 & 0.75 \\
\hline
Averaging CE & 1 & 0.73 & 1.13 \\
\hline
\hline
Channel sounder Tx & 41.1 & 25.6 & 43.5 \\
\hline
Channel sounder Rx 1 Ch corr512 & 61 & 39.4 & 44.7 \\
\hline
Channel sounder Rx 2 Ch corr256 & 70.6 & 45.4 & 49.8 \\
\hline
\end{tabular}
\end{center}
\caption{FPGA resource utilization for Kintex 7 XC7K410T}
\label{util}
\end{table}
}
This channel sounding system is built using USRP X310s\cite{usrpx310}. An USRP X310 contains a Xilinx Kintex 7 FPGA (XC7K410T). FPGA resource utilization numbers for all the 3 CEs and different transmit and receive configurations are shown in Table~\ref{util}. Each USRP X310 contains 2 radio daughter boards, i.e., each X310 has 2 Tx channels and 2 Rx channels. In the channel sounding receiver, either 1 Rx channel or both the Rx channels of an X310 can be used. In case of 1 Rx channel, the FPGA is configured with 1 Rx processing chain as shown in Figure~\ref{rfnoc_rx}. In case of 2 Rx channels, the FPGA is configured with 2 Rx processing chains as shown in Figure~\ref{rfnoc_rx_dual}. In 1 Rx channel configuration, a size 512(can use PN sequences up to length 511) correlator is used. But, due to FPGA resource constraints, in 2 Rx channel configuration, a size 256 correlator is used. Table~\ref{util} shows resource utilization for both single and dual Rx configurations.

\section{Experimental Setup}\label{setup}

\begin{figure}[ht]
\begin{center}
\centerline{\includegraphics[width=\columnwidth]{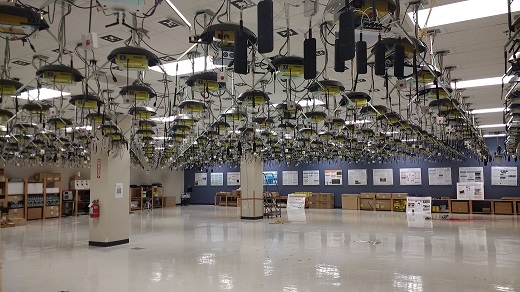}}
\caption{ORBIT Testbed}
\label{orbit}
\end{center}
\end{figure}

\begin{figure}[ht]
\begin{center}
\centerline{\includegraphics[width=\columnwidth]{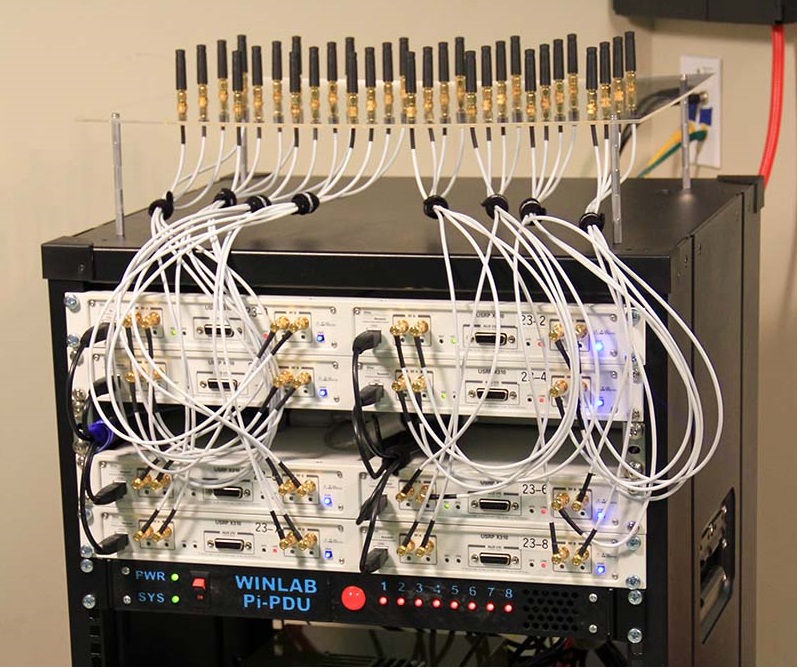}}
\caption{Massive MIMO mini-rack in ORBIT}
\label{mimorack}
\end{center}
\end{figure}

All the channel sounding experiments are conducted in an indoor echoic environment and for that ORBIT (Open-Access Research Testbed) \cite{raychaudhuri2005overview},\cite{orbiturl} is used. ORBIT consists of a grid of wireless nodes each with a Computer, Software Defined Radio (SDR) and an antenna setup. There are a total of 400 nodes in a 20x20 grid as shown in Figure \ref{orbit} and 4 Massive MIMO mini-racks each with 8 USRP X310s \cite{usrpx310} and 32 antennas (16 Tx and 16 Rx) as shown in Figure \ref{mimorack}. Each USRP has 2 RF chains and so 2 Tx and 2 Rx antennas are connected to each USRP. Since multiple receivers are to be used simultaneously, each USRP is given an external Pulse Per Second (PPS) and 10 MHz clock input for time and frequency synchronization respectively.
\begin{figure}[ht]
\begin{center}
\centerline{\includegraphics[width=\columnwidth]{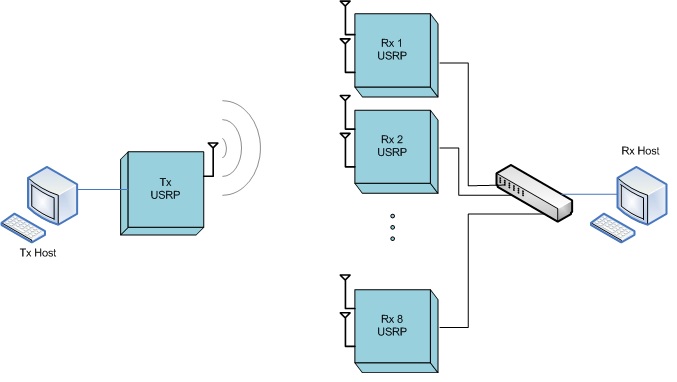}}
\caption{Channel Sounding experiment setup}
\label{HWsetup}
\end{center}
\end{figure}

A single USRP of one of the mini-racks is used as the transmitter and multiple USRPs of a mini-rack located diagonally opposite to the mini-rack with the transmitting USRP are being used as receivers for the experimental setup. Such a setup is used so as to have the longest possible distance between the transmitter and receiver within the Testbed. The transmitting node is controlled by a back-end server and all the receiving nodes are controlled by a different back-end server and the experiment setup looks as shown in Figure~\ref{HWsetup}. Now, the specifications for different parameters such as frequency, sampling rate, number of antennas, of the USRPs and the ranges for those specifications can be seen in \cite{usrpx310}. The specifications that are to be used for this experiment are given in the table below,

\begin{table}
\begin{center}
\begin{tabular}{|l|c|}
\hline
Center Frequency & 5.4 GHz \\
\hline
Sampling Rate & 100 MHz \\
\hline
Number of Transmitting antennas & 1 \\
\hline
Number of Receiving antennas & 16 \\
\hline
PN-Sequence length & 255 \\
\hline
Spreading Factor & 255 \\
\hline
Averaging Factor & 16 \\
\hline
\end{tabular}
\end{center}
\caption{Specifications for experimental setup}
\label{expspecs}
\end{table}

\begin{figure}[ht]
\begin{center}
\centerline{\includegraphics[width=\columnwidth,height= 4 cm]{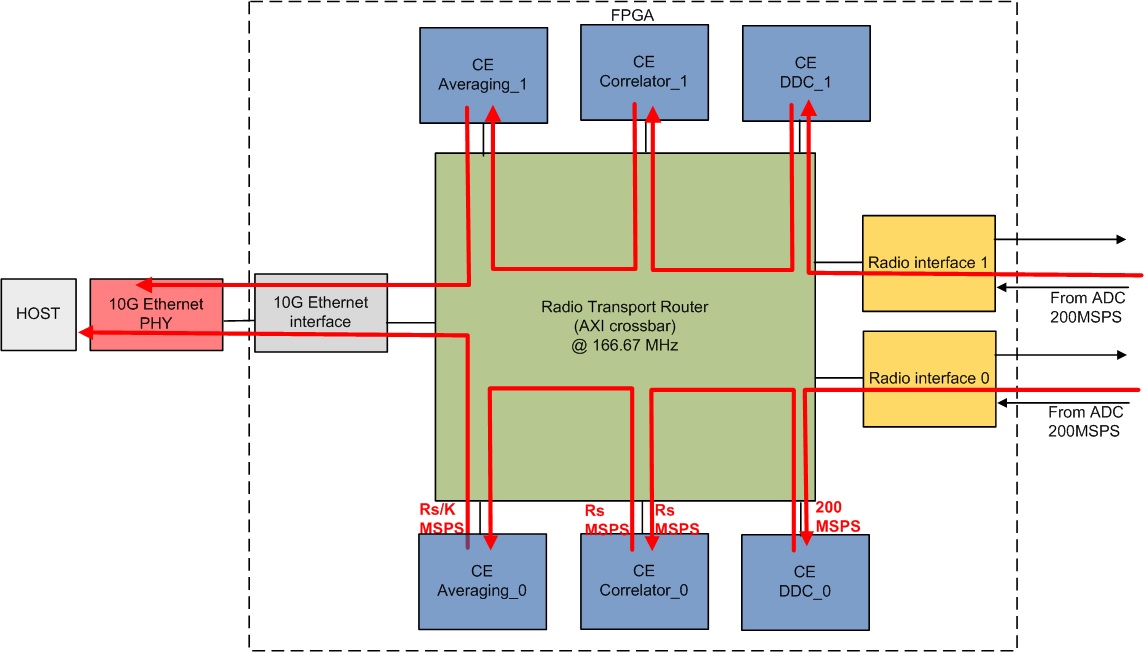}}
\caption{RFNoC Channel Sounding receiver with 2 correlation chains}
\label{rfnoc_rx_dual}
\end{center}
\end{figure}

All the parametric specifications for the experiment we have conducted to show the implementation of the channel sounder are given in Table \ref{expspecs}. The USRPs can be configured to any center frequency between 10 MHz to 6 GHZ, but the frequency selection is based on the antennas being used. We use only Maximal Length Sequences for the experiment but any LFSR based PN sequence up to length 255 can be used in this channel sounder. We use Maximal Length sequences because of the good auto-correlation properties provided which is extremely important for channel sounding. The properties of Maximal length sequences can be seen in \cite{dinan1998spreading}. Spreading Factor is the number of PN sequence bits to use for spreading any data symbol. Here, we spread each incoming symbol with the whole PN sequence and so the Spreading Factor and the Sequence length will always remain equal. Averaging factor is the number of outputs of the correlation block to be averaged. So, if the averaging factor is 16, then 16 outputs of the length of the PN sequence will be averaged.

The following steps are performed on the transmit and receive hosts for the channel sounding experiment.

\textbf{Transmitter}: The transmit host (server) first builds the FPGA  flow graph~\cite{braun2015rfnochost} as shown in Figure~\ref{rfnoc_tx}. Then, the required PN-sequence seed and generator polynomial arguments are sent to the Spreader block in the USRP. Known complex sample is generated at the host and repeatedly sent to  the USRP, which spreads the symbol with the generated PN sequence and the resulting signal is transmitted. The known symbol has magnitude 1 and a certain phase.

\textbf{Receiver}: Host at the receiving end first builds the FPGA flow graph. For this experiment, as both the Rx channels in each X310 are being used, the FPGA is configured as shown in Figure~\ref{rfnoc_rx_dual}. Also, as mentioned in Section~\ref{fpgautil}, due to resource constraints, 2 size 256 correlator CEs are used for this experiment. Then, PN sequence specific arguments for the Correlator block, and averaging factor for the Averaging block (which is the number of correlation sequences to be averaged) are sent to the USRP. Once a "start" signal is sent to the correlator, received signal is correlated and averaged. Output from the Averaging block is then sent to the host for plotting the Power Delay Profile (PDP).

\section{Results}\label{res}

\begin{figure}[H]
\begin{center}
\centerline{\includegraphics[width=\columnwidth,height=5 cm]{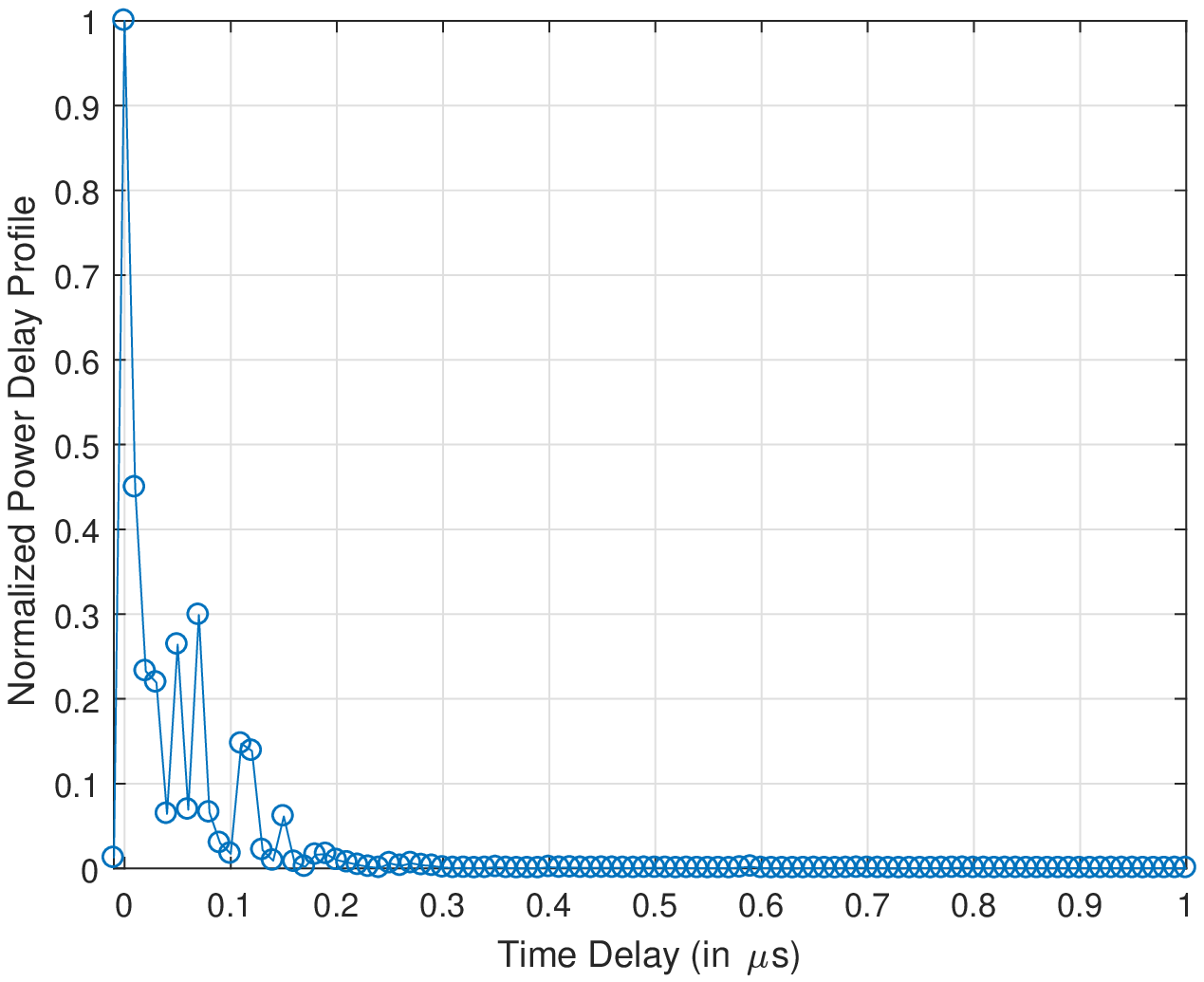}}
\caption{Normalized Power Delay Profile of a single channel}
\label{pdpsinglechan}
\end{center}
\end{figure}

\begin{figure}[H]
\begin{center}
\centerline{\includegraphics[width=\columnwidth,height=5 cm]{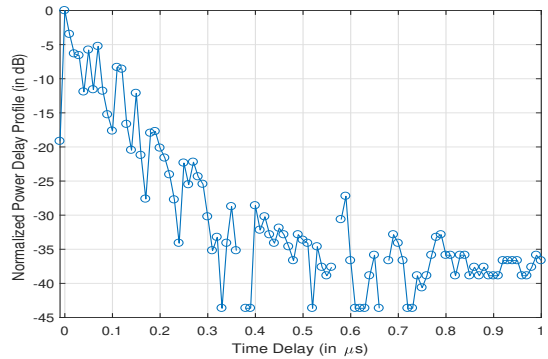}}
\caption{Normalized Power Delay Profile of a single channel (in dB)}
\label{pdpsinglechandb}
\end{center}
\end{figure}

The output of the channel sounder is the Power Delay Profile. So, one binary file per channel is generated consisting of samples with 32-bit resolution. All the binary files are then plotted using either Octave or Matlab. Now, in Figure \ref{pdpsinglechan} and \ref{pdpsinglechandb}, the Normalized Power Delay Profile (PDP), also known as Multipath Intensity Profile, of a single channel can be seen. This is the PDP of one of the 16 channels and is shown as an example of the Channel Sounder output. The Delay Spread and the number of multipath components can be calculated. For example, if the threshold for PDP is set to 0.1 i.e. -10 dB, the delay spread, time difference between first channel tap and the last channel tap above the threshold, comes to approx. $0.12 \ \mu s$. The number of multipath components i.e. the number of channel taps is 8.

\begin{figure}[ht]
\begin{center}
\centerline{\includegraphics[width=\columnwidth]{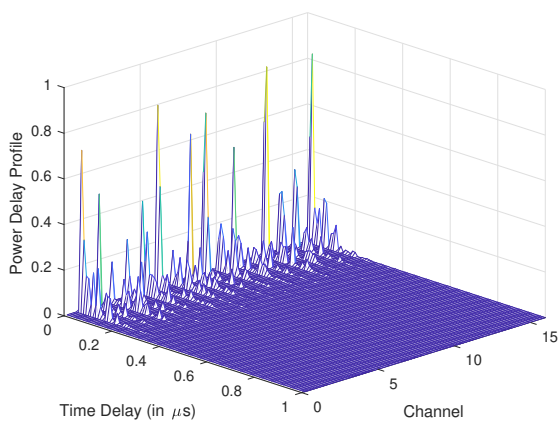}}
\caption{Power Delay Profile of all 16 channels}
\label{pdpallchan}
\end{center}
\end{figure}

\begin{table}[H]
\begin{center}
\begin{tabular}{|c|c|c|}
\hline
Channel & Delay Spread (in $u$s) & Channel Taps \\
\hline
1 & 0.12 & 8 \\
\hline
2 & 0.12 & 4 \\
\hline
3 & 0.22 & 15 \\
\hline
4 & 0.15 & 7 \\
\hline
5 & 0.11 & 8 \\
\hline
6 & 0.01 & 2 \\
\hline
7 & 0.17 & 16 \\
\hline
8 & 0.05 & 4 \\
\hline
9 & 0.02 & 3 \\
\hline
10 & 0.12 & 10 \\
\hline
11 & 0.11 & 6 \\
\hline
12 & 0.16 & 11 \\
\hline
13 & 0.12 & 3 \\
\hline
14 & 0.15 & 10 \\
\hline
15 & 0.11 & 7 \\
\hline
16 & 0.11 & 8 \\
\hline
\end{tabular}
\end{center}
\caption{Delay spread and Channel Taps for Figure \ref{pdpallchan}}
\label{delspr}
\end{table}

Figure \ref{pdpallchan} shows the PDP for all 16 channels. Here, normalized PDP is not shown so that comparison between channels can be done. Also, the signal attenuation for each Transmitter-Receiver pair can be observed. Table \ref{delspr} gives the Delay spread and number of channel taps for each channel. For calculating the parameters in table \ref{delspr}, the PDP for each channel is first normalized and then the threshold for each channel is set to 0.1, the same as that for the single channel example. Normalization is required since the delay spread is calculated with reference to the first channel tap which is the Line-of-sight component and also to keep the threshold value common for all channels.

\section{Conclusion}\label{conc}
Our curiosity and necessity to understand the multipath environment in ORBIT test bed, especially with large scale multi-antenna systems led us to channel sounding experiments. By moving the correlator and averaging blocks to the FPGA, processing time on the host, and amount of data transferred from USRP to host were significantly reduced. This made the experiments much more easier and interesting as results such as receive PDP at different antennas can be observed real time. In this process RFNoC framework was explored, which reduced the system development time and encourages us to move more signal processing tasks to the FPGA. With an addition of few more modules such as frequency synchronization, we hope to use this channel sounding system with mobile nodes as well.  

Source code for the RFNoC blocks and channel sounding application can be found at 
https://github.com/Xilinx/RFNoC-HLS-WINLAB

%We created and implemented a correlator based channel sounder for UWB Large Scale antenna systems. We transmitted a DSSS based sequence and correlated and averaged it at the receivers to obtain an averaged PDP for each receiving antenna. An experiment was conducted using a single transmitter and 16 receivers in an indoor environment. The PDP was plotted, and the Delay spread and number of channel taps were calculated for each antenna

\bibliography{ref}
\bibliographystyle{grcon2016}
\end{document}